\documentclass[twocolumn,prl,aps]{revtex4}
\usepackage{graphicx}

\begin{document}
\title{Mean king's problem with mutually unbiased bases \\
       and orthogonal Latin squares}
\author{A.~Hayashi, M.~Horibe, and T.~Hashimoto}
\address{Department of Applied Physics\\
           Fukui University, Fukui 910-8507, Japan}

\begin{abstract}
The mean king's problem with maximal mutually unbiased bases (MUB's)  
in general dimension $d$ is investigated. 
It is shown that a solution of the problem exists if and only if 
the maximal number ($d+1$) of orthogonal Latin squares exists.
This implies that there is no solution in $d=6$ or $d=10$ dimensions 
even if the maximal number of MUB's exists in these dimensions.
\end{abstract}

\maketitle

\newcommand{\ket}[1]{|\,#1\,\rangle}
\newcommand{\bra}[1]{\langle\,#1\,|}
\newcommand{\braket}[2]{\langle\,#1\,|\,#2\,\rangle}
\newcommand{\bold}[1]{\mbox{\boldmath $#1$}}
\newcommand{\sbold}[1]{\mbox{\boldmath ${\scriptstyle #1}$}}
\newcommand{\tr}[1]{{\rm tr}\left[#1\right]}
\newcommand{\CH}{{\bold{H}}}
\newcommand{\BC}{{\bold{C}}}
\newcommand{\bs}{{\bold{s}}}

\section{Introduction}
In quantum mechanics, one cannot predict the values of two 
or more spin components of a spin-1/2 particle with certainty, 
since the spin operators $\sigma_x$, $\sigma_y$, and $\sigma_z$ 
are not commutable. However, one can ask whether it 
is possible to retrodict the results of a spin 
measurement along more than one possible axis. This question, also known as 
the mean king's problem, was raised and solved by Vaidman, Aharonov, 
and Albert \cite{Vaidman87}. 

More precisely, the problem is formulated as follows.
Alice prepares a spin-1/2 particle in an initial state and gives it to 
Bob. He measures a spin component of the particle along one of the 
$x$, $y$, and $z$ directions and gives the particle back to Alice. 
Without knowledge of Bob's measurement direction and outcome, Alice 
performs some measurement on the particle. Then Bob informs Alice of 
his measurement direction. Alice's task is now to retrodict 
the value of spin component along Bob's measurement direction.

In \cite{Vaidman87} it was shown that Alice can retrodict the values 
of spin component with certainty by utilizing the maximally entangled 
state of two spin-1/2 particles. The quantum-optical version of this 
scheme was also experimentally realized with an average success probability 
of 95.6\% \cite{Schulz03}.

In the king's problem of a spin-1/2 particle, Bob chooses one of the three 
bases, each of which consists of the two eigenstates of $\sigma_i$. 
A natural extension to higher 
dimensions, denoted by $d$, involves $d+1$ mutually unbiased bases 
(MUB's) for Bob's possible measurement bases. 
Here, two orthonormal bases 
$\{\ket{\phi_k}\}_{k=0}^{d-1}$ and $\{\ket{\psi_k}\}_{k=0}^{d-1}$ are 
said to be mutually unbiased 
if $|\braket{\phi_k}{\psi_{k'}}|^2 = 1/d$ for all $k$ and $k'$.
For the king's problem with non-MUB's in $d=2$ dimension, 
see \cite{Menahem89,Horibe05}.

The maximal number of MUB's in $d$ dimension is at most $d+1$.
It is known that if $d$ is a prime or a power of a prime, 
there exists the maximal number $d+1$ of MUB's \cite{Ivanovic81,Wootters89}, 
but this is not known for any other composite numbers.
A maximal set of MUB's plays an important role in quantum 
tomography with Wigner functions in finite dimensions \cite{Gibbons04}.

The king's problem with MUB's has been successfully generalized to higher 
dimensions:
$d=3$ \cite{Aharonov01}, $d=\mbox{prime}$ \cite{Englert01},
and $d=\mbox{power of prime}$ \cite{Aravind03}. 
In these generalizations, it was shown that Alice can retrodict 
the result of Bob's measurement in MUB's if Alice initially prepares a 
$d$ level quantum system as a subsystem of a maximally entangled composite 
system and in the end she performs a projection 
measurement in an appropriate basis on the composite system.

A natural question is then whether the king's problem with MUB's can always 
be solved in the same way provided that the maximal number of MUB's 
exists. In this paper, we will study this problem and show that the 
existence of a solution of the king's problem with MUB's is equivalent to
that of the maximal number of mutually orthogonal Latin squares. 
A possible relation between MUB's and orthogonal Latin squares was 
conjectured by Wootters \cite{Wootters04}.

Combined with known results on orthogonal Latin squares, 
our result first gives an alternative proof of the existence of solutions 
of the king's problem in prime power dimensions, and secondly it also implies 
that there is no solution in $d=6$ and $d=10$ dimensions even if the 
maximal set of MUB's can be constructed in these dimensions.  

\section{King's problem with mutually unbiased bases} 
The problem we will consider is the following. In a $d$-dimensional complex 
vector space $\BC^d$, we consider $d+1$ orthonormal bases 
labeled by $A(=0,1,\ldots,d)$.  By $\ket{A,a}\ (a=0,1,\ldots,d-1)$, 
we denote a state vector in base $A$. 
We assume that the bases are mutually unbiased:
\begin{eqnarray}
   \left| \braket{A,a}{A',a'} \right|^2 =
   \delta_{AA'}\delta_{aa'} + (1-\delta_{AA'})\frac{1}{d}.
                        \label{MUB_condition}
\end{eqnarray}
Alice and Bob share a maximally entangled state on $\BC^d \otimes \BC^d$:
\begin{eqnarray}
    \ket{\Phi} = \frac{1}{\sqrt{d}}\sum_{k=0}^{d-1} \ket{k}\otimes\ket{k},
\end{eqnarray}
where $\{\ket{k}\}_{k=0}^{d-1}$ is a reference orthonormal base in $\BC^d$.
In terms of base $A$, the maximally entangled state $\ket{\Phi}$ is 
written as
\begin{eqnarray}
  \ket{\Phi} = \frac{1}{\sqrt{d}}\sum_{a=0}^{d-1} 
            \ket{\overline{A,a}}\otimes\ket{A,a},
\end{eqnarray}
where an overlined state $\ket{\overline{\phi}}$ for a state $\ket{\phi}$ 
is defined as $\ket{\overline{\phi}}=\sum_k \ket{k}\braket{k}{\phi}^*$ 
by the use of the reference base.

Bob randomly chooses base $A \in \{0,1,\ldots,d\}$ and measures his local 
system, the second component of the tensor product, 
and obtains the outcome $a \in \{0,1,\ldots,d-1\}$.
The post-measurement state is then given by
\begin{eqnarray}
  \ket{\Phi_{A,a}} \equiv \ket{\overline{A,a}}\otimes\ket{A,a}.
\end{eqnarray}
Here we note that the inner product of $\ket{\Phi_{A,a}}$'s is calculated 
as
\begin{eqnarray}
\braket{\Phi_{A,a}}{\Phi_{A',a'}}=\left| \braket{A,a}{A',a'} \right|^2,
                         \label{inner_product}
\end{eqnarray} 
where the right-hand side is given in Eq.(\ref{MUB_condition}).
Without knowing Bob's measurement base $A$ and outcome $a$, 
Alice performs a projective measurement on the composite system in the base 
$\{\ket{I}\}_{I=0}^{d^2-1}$ of $\BC^d \otimes \BC^d$.
After the measurement, Alice is informed of Bob's measurement base $A$. 
Alice's task is now to estimate Bob's outcome $a$ from her measurement outcome
$I$ and Bob's base $A$. Let us write Alice's estimate for 
Bob's outcome as $s(I,A) \in \{0,1,\ldots,d-1\}$.

It is clear that Alice's success probability is 1 if and only if 
the following conditions are satisfied:
\begin{eqnarray}
    \braket{\Phi_{A,a}}{I}=0\ \mbox{for}\ a \ne s(I,A).
                                 \label{p1_condition}
\end{eqnarray}
In the following, we will study the above conditions in detail. 

First we show that the set $\Phi$ consisting of $d(d+1)$ states 
$\{\ket{\Phi_{A,a}}\}_{A=0,a=0}^{A=d,a=d-1}$ 
is complete in the composite space $\BC^d \otimes \BC^d$. Suppose a linear relation with some coefficients 
$c_{A,a}$ holds,
\begin{eqnarray}
      \sum_{A,a} c_{A,a} \ket{\Phi_{A,a}} = 0.  \label{linear_relation}
\end{eqnarray}
Multiplying bra vector $\bra{\Phi_{A',a'}}$ from the left and using the 
relation (\ref{inner_product}), 
we immediately find 
\begin{eqnarray}
  c_{A',a'}+\sum_{A(\ne A')} \frac{1}{d}\sum_a c_{A,a} =0,
\end{eqnarray}
which implies that the coefficient $c_{A,a}$ should be independent 
of $a$. Now consider the subset $\Phi'$ which consists of $d^2$ states 
obtained by removing $d$ states $\{\ket{\Phi_{A,a}}\}_{A \ne 0, a=0}$ 
from $\Phi$. 
Assume a linear relation of type Eq.(\ref{linear_relation}) for 
the states in $\Phi'$. We find that $c_{A,a}=0$ for $A \ne 0$ and any $a$, 
since $c_{A,0}=0$ for $A \ne 0$ by assumption. 
The linear relation is then reduced to
\begin{eqnarray}
 \sum_{a=0}^{d-1} c_{0,a}\ket{\Phi_{0,a}} = 
 c_0 \sum_{a=0}^{d-1} \ket{\Phi_{0,a}} = \sqrt{d}\,c_0 \ket{\Phi}=0,
\end{eqnarray}
implying $c_{0,a}(\equiv c_0)=0$. 
Thus the $d^2$ states in $\Phi'$ are linearly 
independent and the set $\Phi$ is complete in the $d^2$-dimensional composite 
space $\BC^d \otimes \BC^d$.

Next we consider another subset $\Phi''_I$ obtained by removing $d+1$ states 
$\{\ket{\Phi_{A,s(I,A)}}\}_{A=0}^d$ from 
$\Phi$ for a given $I$. It is easy to see that the $d^2-1$ 
states in $\Phi''_I$ are linearly independent and span a $d^2-1$ subspace of 
$\BC^d \otimes \BC^d$, since a linear relation of type 
Eq.(\ref{linear_relation}) with $c_{A,s(I,A)}=0$ requires that all 
coefficients $c_{A,a}$'s be zero.

Now we go back to the conditions (\ref{p1_condition}) for Alice's measurement 
base $\{\ket{I}\}_{I=0}^{d^2-1}$ and estimate $s(I,A)$.
The conditions (\ref{p1_condition}) require that each state $\ket{I}$ 
should be orthogonal to the $d^2-1$ dimensional subspace spanned by 
$\Phi''_I$ and can be uniquely determined up to an irrelevant phase factor.

We find that state $\ket{I}$ satisfying conditions (\ref{p1_condition})
is given by
\begin{eqnarray}
  \ket{I} = \frac{1}{\sqrt{d}}\sum_{A=0}^d \ket{\Phi_{A,s(I,A)}} -\ket{\Phi}.
                     \label{I_definition}
\end{eqnarray}
It is readily verified that the state $\ket{I}$ satisfies the conditions 
(\ref{p1_condition}) as
\begin{eqnarray}
  \braket{\Phi_{A,a}}{I} = \frac{1}{\sqrt{d}}\delta_{a,s(I,A)},
                    \label{I_condition}
\end{eqnarray}
and $\ket{I}$ is normalized as $\braket{I}{I}=1$.

It remains to determine the condition for Alice's estimate $s(I,A)$ under 
which $\{\ket{I}\}_{I=0}^{d^2-1}$ is an orthonormal base in 
$\BC^d \otimes \BC^d$. Namely, $d^2$ states defined in Eq.(\ref{I_definition})
should be orthogonal to each other. The inner product $\braket{I}{I'}$ can be 
easily calculated and we find the orthogonality holds if and only if 
\begin{eqnarray}
    \sum_{A=0}^{d} \delta_{s(I,A),s(I',A)} = 1\ \mbox{for}\ I \ne I'.
                               \label{S_condition}
\end{eqnarray}
This is a necessary and sufficient condition for Alice's success 
probability to be equal to 1.

We can interpret the conditions (\ref{S_condition}) in terms of a set of 
character strings which satisfies a certain relation. Consider a character 
string $\bs$ of length $d+1$ with each character chosen from the set 
$\{0,1,\ldots,d-1\}$:
\begin{eqnarray}
     \bs = s_0s_1s_2 \cdots s_d,\ \ s_A \in \{0,1,\ldots,d-1\}.
\end{eqnarray} 
In this paper, two strings $\bs$ and $\bs'$ of this type are said to be 
"mutually unbiased" if they share a common character at exactly one 
common place,
\begin{eqnarray}
    \sum_{A=0}^{d} \delta_{s_A,s_A'} = 1.
\end{eqnarray}

With this definition, the necessary and sufficient condition 
(\ref{S_condition}) is 
equivalent to the existence of $d^2$ mutually unbiased strings of 
length $d+1$ of $d$ kinds of characters. 
In Fig.1 we give an example of a set of nine mutually unbiased strings of 
length 4 in the case of $d=3$.
The question is for what $d$ the set of unbiased strings of this kind can 
be accommodated. 
Before answering this question, however, we will study another form of the 
necessary and sufficient condition on $s(I,A)$.

As mentioned before, $d^2$ normalized states $\ket{I}$'s defined by 
Eq.(\ref{I_definition}) should form an orthonormal base. 
From the orthogonality, we obtained the condition (\ref{S_condition}). 
Here we study the condition derived from the completeness 
$\sum_{I=0}^{d^2-1} \ket{I}\bra{I} = \bold{1}$, which is equivalent to the 
orthogonality. Since the set of states 
$\{\ket{\Phi_{A,a}}\}_{A=0,a=0}^{A=d,a=d-1}$ is complete, the completeness 
of $\{\ket{I}\}_{I=0}^{d^2-1}$ can be expressed as
\begin{eqnarray}
  \sum_{I=0}^{d^2-1} \braket{\Phi_{A,a}}{I}\braket{I}{\Phi_{A',a'}}
  = \braket{\Phi_{A,a}}{\Phi_{A',a'}}.
\end{eqnarray}
Using Eq.(\ref{I_condition}) for the left-hand side and 
Eqs.(\ref{inner_product}) and (\ref{MUB_condition}) for the right-hand side,
we obtain
\begin{eqnarray}
  \sum_{I=0}^{d^2-1} \delta_{a,s(I,A)}\,\delta_{a',s(I,A')}
  = d\,\delta_{A,A'}\delta_{a,a'}+(1-\delta_{A,A'}),
                \label{L_condition}
\end{eqnarray}
which is also one form of a necessary and sufficient condition for Alice to
guess Bob's outcome with certainty.

The condition (\ref{L_condition}) turns out to be that of the existence of 
$d+1$ orthogonal $d$ by $d$ Latin squares . First we introduce the 
definition of orthogonal Latin squares formulated by 
Wootters \cite{Wootters04}.

Consider a collection of $d^2$ points, which can be regarded as points of 
a $d \times d$ square lattice. They are divided into $d$ groups so that each group consists of $d$ points and any point belongs to only one group. 
A way of partitioning of this kind is called "striation" or $d \times d$ 
Latin square. 
Two Latin squares are said to be orthogonal if each group of either 
Latin square has exactly one point in common with each group in the 
other.

Let us label a point by $I(=0,1,\ldots,d^2-1)$, a group by 
$a(=0,1,\ldots,d-1)$, and a Latin square by $A$. 
Then a Latin square $A$ is uniquely characterized by its "striation function" 
$s(I,A)$. 
Namely, $a=s(I,A)$ means that point $I$ in Latin square $A$ belongs to 
group $a$.

Now it is easy to see that the condition (\ref{L_condition}) can be 
satisfied by some estimate $s(I,A)$ if
and only if there exist $d+1$ orthogonal $d \times d$ Latin squares.
The if part is clear since the left-hand side of Eq.(\ref{L_condition})
counts the number of common points belonging to group $a$ in Latin 
square $A$ and group $a'$ in Latin square $A'$. On the other hand, 
summing over $a'$ in Eq.(\ref{L_condition}), we obtain 
$\sum_{I=0}^{d^2-1}\delta_{a,s(I,A)} = d$. This means estimate $s(I,A)$ 
is a striation function and Eq.(\ref{L_condition}) implies that 
two Latin squares $A$ and $A'$ are orthogonal if $A \ne A'$. 
An example of a set of 4 orthogonal $3 \times 3$ Latin squares is 
given in Fig.1.

\begin{figure}
\includegraphics[width=8cm]{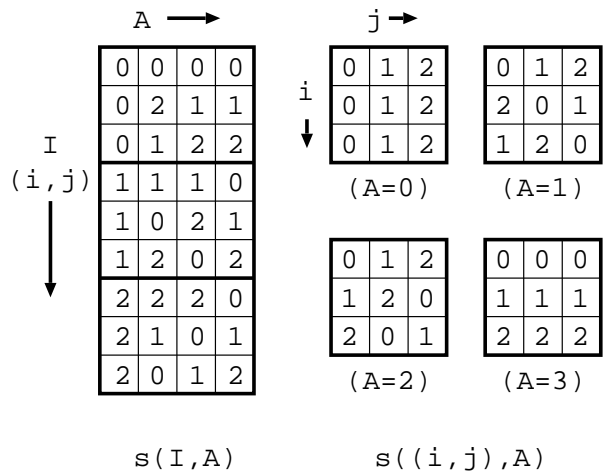}
\caption{\label{}
Example of nine mutually unbiased strings and four orthogonal
$3 \times 3$ Latin squares. On the left, strings $s(I,A)$ as a $9 \times 4$ 
matrix are shown. 
When each row is regarded as a string of length 4, any pair of 
nine strings shares a character at exactly one common place. 
On the right, four 
mutually orthogonal $3 \times 3$ Latin squares are shown, which are 
constructed by the same $s(I,A)$ interpreted as a striation function.
In this example, $s((i,j),A)$ is taken to be $j-Ai\hspace{-0.8em}\pmod{3}$ 
for $A=0,1,2$ and $i$ for $A=3$. The site $(i,j)$ in a square is numbered by 
$I(=3j+i)$.}
\end{figure}

Some known results on the maximum number $M(d)$ of $d \times d$ mutually 
orthogonal Latin squares are summarized in \cite{Wootters04} as follows:
\begin{itemize}
\item For any $d$, $M(d) \le d+1$.
\item If $d$ is a power of a prime, $M(d)=d+1$.
\item $M(6)=3$.
\item If $d-1$ or $d-2$ is divisible by four, and if $d$ is not the sum 
of the squares of two integers, then $M(d) < d+1$.
\item $M(10) < 11$.
\end{itemize}

If $d$ is a power of a prime, there exists a maximal set of MUB's  
in $\BC^d$ \cite{Wootters89} and there exist maximal $d+1$ orthogonal 
$d\times d$ Latin squares. 
The latter can be seen by explicitly constructing them with the striation 
function 
\begin{eqnarray}
    s((i,j),A) &=& \left\{  \begin{array}{ll}
                               j-Ai & (A=0,1,\cdots,d-1) \\
                               i    & (A=d)              \\
                          \end{array}
                   \right. , \nonumber \\
               & & \hspace{-3em} (i,j=0,1,\cdots,d-1),\ (I=jd+i),
\end{eqnarray}
where $j-Ai$ on the right-hand side should be calculated under  
the rules of the field $F(d)$, which are reduced to the ordinary calculus 
modulo $d$ when $d$ is a prime. See Fig.1 for an example in the case of $d=3$.
Therefore, the mean king's problem of the type 
considered in this paper has a solution. 
This is what Aravind has already shown explicitly \cite{Aravind03}, and 
the present paper provides an alternative proof. 

When $d$ is not a power of a prime, it is in general an open question 
how many MUB's exist in $\BC^d$. 
However, we know that there does 
not exist maximal $d+1$ orthogonal $d \times d$ Latin squares for $d=6$ or 
$d=10$. Thus we conclude that there is no solution to the king's problem 
of the type considered in this paper in these dimensions, even if there exists 
a maximal set of MUB's.

We have obtained two forms of conditions: the existence of 
$d^2$ mutually unbiased strings of length $d+1$ and 
$d+1$ orthogonal $d \times d$ Latin squares.
Both conditions are necessary and sufficient 
conditions for the existence of a solution of the king's problem.
This implies that the two conditions are equivalent to each other.
In the next section, however, we will directly show the equivalence 
without resorting to the king's problem. 

\section{Mutually unbiased strings and orthogonal Latin squares}
As shown in the preceding section,
the following relations are the existence conditions of $d^2$ mutually 
unbiased strings of length $d+1$ composed of a set of $d$ characters,
\begin{eqnarray}
  && \sum_{A=0}^d\delta_{s(I,A),s(I',A)}
    = (d+1)\delta_{I,I'} + (1-\delta_{I,I'}),\nonumber \\
  && \hspace{8em} (I,I'=0,1,\ldots,d^2-1),
                 \label{SS_condition}
\end{eqnarray}
where $s(I,A) \in \{0,1,\ldots,d-1\}$ should be regarded as a character 
at the $A$th place of string $\bs_I$.
On the other hand, the existence of $d+1$ mutually orthogonal Latin 
squares implies 
\begin{eqnarray}
 && \sum_{I=0}^{d^2-1} \delta_{a,s(I,A)}\,\delta_{a',s(I,A')}
  = d\,\delta_{A,A'}\delta_{a,a'}+(1-\delta_{A,A'}), \nonumber \\
 && \hspace{9em} (A,A'=0,1,\ldots,d), \nonumber \\
 && \hspace{9em} (a,a'=0,1,\ldots,d-1),
                \label{LL_condition}
\end{eqnarray}
where $s(I,A)$ should be interpreted as a striation function, meaning 
point $I$ in Latin square $A$ belongs to group $a=s(I,A)$. 
We will show that the two conditions (\ref{SS_condition}) and 
(\ref{LL_condition}) are in fact equivalent as expected from the argument 
in the preceding section.

We will show the equivalence by explicitly "diagonalizing" the two conditions.
First we define the Fourier transform of $\delta_{a,s(I,A)}$ as
\begin{eqnarray}
   u_{A,\alpha:I} \equiv \frac{1}{d}\sum_{a=0}^{d-1} \omega^{\alpha a}
                    \delta_{a,s(I,A)},
\end{eqnarray}
where $\omega$ is a primitive $d$th root of unity. We note that 
$u_{A,0:I}$ is given by $1/d$, independent of $A$ and $I$. 
In terms of the Fourier transform $u$, the two conditions 
(\ref{SS_condition}) and (\ref{LL_condition}) are then rewritten as
\begin{eqnarray}
 \sum_{A=0}^{d}\sum_{\alpha=0}^{d-1}\, u_{A,\alpha:I}^*\, u_{A,\alpha:I'} 
     = \delta_{I,I'} + \frac{1}{d}
             \label{SSS_condition}
\end{eqnarray}
and
\begin{eqnarray}
 && \sum_{I=0}^{d^2-1}\, u_{A,\alpha:I}\,u_{A',\alpha':I}^* \nonumber \\
 && \hspace{3em}   =  \delta_{A,A'}\delta_{\alpha,\alpha'}
                     +\delta_{\alpha,0}\delta_{\alpha',0}(1-\delta_{A,A'}),
              \label{LLL_condition}
\end{eqnarray}
respectively.

Recalling $u_{A,0:I}=1/d$, we introduce a $d^2 \times d^2$ matrix $U_{J,I}$
in the following way. Here the subscript $J$ collectively represents either 
one of $d^2-1$ pairs $(A,\alpha \ne 0)$ or pair $(A=0,\alpha=0)$.
Let us define the matrix $U$ as
\begin{eqnarray}
  U_{J,I} = \left\{ \begin{array}{ll}
                        u_{A,\alpha:I}  & \mbox{if}\ J=(A,\alpha \ne 0) \\
                        u_{0,0:I}       & \mbox{if}\ J=(0,0). \\
                    \end{array} 
            \right.
\end{eqnarray} 

The two conditions (\ref{SSS_condition}) and (\ref{LLL_condition}) turn 
out to be $U^+U=\bold{1}$ and $UU^+=\bold{1}$, respectively, 
which are evidently equivalent to each other.
We thus conclude that a set of $d^2$ mutually unbiased strings with length 
$d+1$ is equivalent to the existence of $d+1$ mutually orthogonal 
$d \times d$ Latin squares. 

\section{Concluding remarks}
We have shown that a solution of the king's problem with MUB's 
requires the maximal number of orthogonal Latin squares. Therefore, 
we can conclude that there is no solution to the problem in some composite 
number of dimensions (6 and 10) for which the maximal set of orthogonal 
Latin squares is known not to exist.

One can, however, ask whether there is a solution to the king's 
problem with non-MUB's. Let us impose three restrictions on the problem: 
the initial state prepared by Alice is a maximally entangled state, 
Alice can perform only a projective measurement (not POVM), and 
Bob's set of measurement bases is complete. 
By completeness of a set of bases, we mean an unknown state is completely 
determined by repeated measurements in those bases, 
if infinitely many copies of the state are available. 

In the case of $d=2$, Horibe {\it et al}.\ \cite{Horibe05} studied the king's 
problem with non-MUB's (bases given by three nonorthogonal spin directions) 
by allowing Alice to perform POVM measurement.
It was found that a solution to the problem with the three restrictions 
exists only if the three spin directions are orthogonal (MUB's).
For general dimensions, we conjecture that the king's problem with the 
three restrictions has a solution only if the set of bases is the maximal 
set of MUB's.

It should be noted that we can construct a solution to the king's problem 
with non-MUB's if the completeness condition of the bases is omitted. 
Assume that a composite dimension $d$ is factored as $d_1d_2$ by two prime 
powers. 
Factoring the $d$-dimensional space into two subsystems with dimension 
$d_1$ and $d_2$, we have a solution of the king's problem with $d_i+1$ MUB's 
for each subsystem $i=1,2$. 
Now take $d+1$ bases on the composite system out 
of $(d_1+1)(d_2+1)$ bases given by tensor products of the two MUB's 
of subsystems. These $d+1$ bases are not in general MUB's on the composite 
system. But it is clear that there is a solution to the king's problem with 
these bases.

\end{document}